\documentclass[twocolumn,showpacs,prb,amsmath,amssymb]{revtex4}
\usepackage{graphicx}
\usepackage{bm}
\DeclareMathOperator{\Tr}{Tr}
\begin{document}
\title{Theory of nuclear induced spectral diffusion: Spin decoherence
of phosphorus donors in Si and GaAs quantum dots}    \author{Rogerio
de Sousa} \author{S. \surname{Das Sarma}} \affiliation{ Condensed
Matter Theory Center, Department of Physics, University of Maryland,
College Park, MD 20742-4111}  \date{\today}
\begin{abstract}
We propose a model for spectral diffusion of localized spins  in
semiconductors due to the dipolar fluctuations of lattice nuclear
spins.  Each nuclear spin flip-flop is assumed to be independent, the
rate for this process being calculated by a method of moments.  Our
calculated spin decoherence time $T_{M}=0.64$ ms for donor electron
spins in Si:P is  a factor of two longer than spin echo decay
measurements. For $^{31}$P  nuclear spins we show that spectral
diffusion is well into the motional  narrowing regime. The calculation
for GaAs quantum dots gives $T_{M}=10-50$ $\mu$s depending on the
quantum dot size.  Our theory indicates that nuclear induced spectral
diffusion should not  be a serious problem in developing spin-based
semiconductor quantum computer architectures.
\end{abstract}
\maketitle
%

\section{Introduction}

Electron and nuclear spins in semiconductors are promising  qubit
candidates for quantum computation  because their intrinsic quantum
two level nature together with existing semiconductor microelectronics
technology  can potentially satisfy the strict control and scalability
requirements of a quantum computer (QC). Hence electron spins in
quantum dots (QD)\cite{loss98} and donor impurities\cite{vrijen99} as
well as nuclear spins in  semiconductors\cite{kane98} have been
suggested as candidate building  blocks for feasible QC
architectures.\cite{dassarma01} However, to build such a device major
advances in single spin manipulations are needed, and for this purpose
realistic calculations of semiconductor spin dynamics are essential to
guide the experimental effort currently taking place.  A question of
particular importance is whether a localized spin will remain
unaffected by the many interactions invariably present in a
semiconductor environment during a time interval long enough for fault
tolerant quantum computation (equivalent to $10^{4}-10^{6}$ quantum
gating times.\cite{preskill98})  In a recent paper\cite{desousa02} we
showed that spin coherence of bound electronic states in
semiconductors is limited by spin-spin interactions at low
temperatures.  When this interaction is between the qubits themselves,
it can in principle be incorporated into the QC Hamiltonian, although
this will lead to more complicated gate sequences.  In particular we
are not aware of any theoretical QC work specifically working out such
gate sequences incorporating inter-qubit interactions.  Therefore it
is instructive to analyze the error introduced by ignoring some of
these interactions,  as we did in the case of dipolar coupled spin
qubits.\cite{desousa02} The presence of many non-resonant spins in the
system, such as lattice nuclei, also leads to phase fluctuation of the
spin qubit, an effect which is hard to control. This has been denoted
Spectral Diffusion (SD) since the qubit Zeeman frequency will diffuse
through the spin resonance line.  Spectral diffusion specifically
refers to fluctuations  in the Zeeman frequency $\gamma
B_{\textrm{eff}}=g \mu_{B} B_{\textrm{eff}}/\hbar$ (where $g$ is the
effective $g$-factor, $\mu_{B}=e\hbar/2mc$ the Bohr magneton, and
$B_{\textrm{eff}}$ the effective local magnetic field) of the electron
due to external effects arising from the semiconductor
environment. Note that such fluctuations could arise either because
the effective magnetic field $B_{\textrm{eff}}$ is changing
dynamically or because the electron $g$-factor is varying. There are
many physical processes leading to spectral diffusion, and here we are
specifically interested in the limiting processes at the lowest
temperatures.  The physical process of interest to us is therefore
dipolar  nuclear fluctuations.

The first order of magnitude estimate of this effect was applied to
Si:P donor electrons,\cite{chiba72} while our recent
paper\cite{desousa02} used the same methods to estimate the SD rate in
GaAs QDs. However, these estimates assumed {\it a priori} that nuclear
pairs flip-flopped slowly (and hence the echo decay was $\sim
\exp-\tau^{3}$) with a rate given by an unjustified phenomenological
equation [Eq. (15) in Ref.~\onlinecite{chiba72} and Eq. (8) in
Ref.~\onlinecite{desousa02}].  Here we propose a new description for
this decoherence mechanism, arising from dipolar fluctuations of the
lattice nuclear spins, affecting the qubit Zeeman frequency through
hyperfine coupling. Even though we still treat each nuclear pair as an
independent Markovian random variable (an approximation which seems
reasonable for temperatures well above nuclear dipolar ordering,
happening on the nanokelvin scale), our new theory describes fast and
slow flip-flops on the same footing incorporating motional narrowing
effects previously absent in former treatments (which happens when the
fluctuation is so fast that SD is suppressed). We also derive
microscopic expressions for these flip-flop rates, leading to a more
refined calculation of nuclear SD for GaAs QDs and Si:P donor
electrons, together with the first treatment of this effect for a
$^{31}$P donor nucleus.

In the case of localized spins precessing in a magnetic field $B$,
knowledge of three phenomenological parameters is sufficient to
describe the spin $1/2$ dynamics: The gyromagnetic ratio $\gamma$
which determines the precessing frequency (or equivalently the $g$
factor, $g = 2mc\gamma/e$ with $e$ the electronic charge, $m$ the bare
electron mass and $c$ the speed of light), the longitudinal relaxation
time or spin-flip time $T_{1}$, and finally the transverse relaxation
time or dephasing time $T_{2}$, which is often denoted coherence time
since it sets the time scale of coherent superpositions between states
along the B field direction.\cite{hu01} However, electron spins in a
solid state environment often have different precession frequencies,
either due to hyperfine fields from nearby nuclear spins or from
unequal gyromagnetic ratios (arising, for example, from varying
carrier effective mass).  Therefore the transverse magnetization of a
spin ensemble will decay in a time scale $T_{2}^{*}$ which is in
general much shorter than  the single spin dephasing time $T_{2}$.
The latter time scale can be measured using a $\pi/2-\pi$ spin echo
sequence.\cite{slichter_abragam} The time it takes for this echo to
decay to $1/e$ of its initial value conveniently defines our single
spin coherence time and has been historically called $T_{M}$ (spin
memory time)\cite{mims72} since the echo envelope usually does not
decay exponentially as predicted by the Bloch equations from which
$T_{2}$ was first defined\cite{hu01} (In appendix B we show that
measuring a $\pi/2-\pi$ echo is equivalent to measuring the modulus
squared of a single spin off diagonal density matrix element).

It has been known for a very long time that SD caused by nearby
non-resonant spins is usually the dominant echo decay
mechanism.\cite{herzog56} However, all former SD theories assumed a
{\it single} relaxation rate for the non-resonant spins, an
approximation perfectly suitable for ``$T_{1}$ samples'', whereby
these spins change their states through spin-flips only.  The theories
of B. Herzog and E.L. Hahn,\cite{herzog56} and later J.R. Klauder and
P.W. Anderson,\cite{klauder62} described the central spin Zeeman
frequency as a random variable evolving in time according to Gaussian
and Lorentzian conditional probabilities respectively. These
assumptions lead to a $\pi/2-\pi$ echo decay of the form
$\exp{(-2T_{1}^{-1}\delta^{2}\tau^{3}/3)}$ and
$\exp{(-T_{1}^{-1}\delta\tau^{2})}$ respectively as long as $\tau \ll
T_{1}$, with $2\tau$ being the time interval between the first pulse
and the echo. Both $\delta$ ($t\rightarrow \infty$ linewidth for the
conditional probabilities) and $T_{1}$ (spin-flip time of non-resonant
spins) are parameters that can in principle be calculated from the
interactions. If the condition $\tau \gg T_{1}$ is satisfied, one
obtains $\exp{(-T_{1}\delta^{2} \tau)}$ and $\exp{(-\delta \tau)}$
respectively, characterizing the motional narrowing regime.
Interestingly, Gaussian SD correctly describes motional narrowing
since $T_{1}$ appears in the numerator, but Lorentzian SD does not,
the decay being independent of $T_{1}$. Motivated by this inadequacy
of the Lorentzian theory, G.M. Zhidomirov and K.M.
Salikhov\cite{zhidomirov69} proposed a many parameter model, which
treated the number of flips of a spin $i$ during a time interval $t$
as a Poisson random variable parametrized by $t/T_{1}$, the frequency
change on the central spin being $\Delta_{i}$. Their theory obtained
the correct motional narrowing limit and agreed with experiment in
dilute $T_{1}$ samples where the non-resonant spins are randomly
distributed. Our problem, however, is in a completely different
regime. Here SD is caused by Si and GaAs lattice nuclei, which have
$T_{1}$ of the order of hours and hence the relevant time scale is
given by the dipolar interaction, which varies substantially depending
on the specific pair flip-flopping (such a system is denominated a
``$T_{2}$ sample'', since $T_{2}\ll T_{1}$ for the spins that create
the SD effect). We generalize the latter theory\cite{zhidomirov69} to
many relaxation rates $T_{nm}^{-1}$, each corresponding to a pair
$n,m$ of nuclear spins.  We also present a microscopic theory to
calculate these flip-flop rates. Altogether this approach is, to our
knowledge, the first systematic attempt to describe SD in $T_{2}$
samples.  We also show that our theory reduces to the earlier simple
approximations of Refs~\onlinecite{desousa02,klauder62} in the
appropriate limits.

The phosphorus donor impurity in silicon is the textbook example of a
localized electron spin in a semiconductor. It has been extensively
studied experimentally using Electron Spin Resonance
(ESR)\cite{feher59a,feher59b,feher61} and successful theories for its
gyromagnetic ratio and $T_{1}$ were
developed.\cite{feher61,haseg_roth60} However $T_{M}$ for P in Si
remained unexplained, even though it was measured thirty years
ago.\cite{chiba72} Our model leads to a $T_{M}$ two times longer than
the measured value (this agreement should be considered reasonable
since existing theories for $T_{1}$ are also off by a factor of
two.\cite{feher61,haseg_roth60}) Our theory predicts a smooth
transition of the echo envelope from Gaussian SD to motional narrowing
behavior, this transition being well described by a correlation
function.  If the echo decay of $^{31}$P donor nuclei is measured, we
predict an echo envelope purely exponential, well into this motional
narrowing regime, with $T_{M}=0.60$ s.  An important point discussed
here is how SD is rapidly suppressed by reducing the amount of nuclear
magnetic moments in the lattice. Our calculations show that isotopic
purification of Si (exchanging spin-$1/2$ $^{29}$Si nuclei by spin-$0$
$^{28}$Si) may lead to coherence times as long as $100$ ms for P
impurities in Si, a result supported by recent
experiments.\cite{tyryshkin03} Unfortunately, Ga and As nuclei have no
stable spin-$0$ isotopes, hence it seems that the only way to increase
spin coherence in these materials is to suppress flip-flop events by
nuclear polarization, as can, for example, be done by applying a
strong external magnetic field or by using the Overhauser effect.
Furthermore, we recently reported the first $T_{M}$ calculations for a
GaAs quantum dot.\cite{desousa02} This was particularly important
since $T_{M}$ has never been measured in this system, and a realistic
assessment of the feasibility of a quantum dot quantum computer was
needed.  The detailed calculation presented here confirms our previous
estimates.  Hence the present paper together with other recent
calculations of g factor\cite{kiselev98} and $T_{1}$\cite{khaetskii01}
available in the quantum dot literature provides a general picture for
electron spin dynamics in these heterostructures.

It is instructive to clarify the relationship between our results and
recently published theories\cite{khaetskii02} on related issues.  In
Ref.~\onlinecite{khaetskii02} the authors considered a Hamiltonian
which contained only hyperfine couplings between a single electron and
the lattice nuclei, discarding the essential ingredient of the
spectral diffusion effect, which is the dipolar interaction between
nuclei.  Hence their mechanism is based on flip-flops between electron
and nuclear spins.  But when a $B$ field is applied electron-nuclear
flip-flops are forbidden by energy conservation, since the nuclear
Zeeman energy is $10^{3}$ times smaller than the electronic Zeeman
splitting.  Therefore their mechanism is only relevant at low $B$
fields, when the hyperfine coupling is of the same magnitude or
greater than the electronic Zeeman energy, leading to the condition
$B\lesssim \hbar \gamma_{I}|\Psi(0)|^{2}\ll 100$ G where $\gamma_{I}$
is the nuclear gyromagnetic ratio and
$|\Psi(0)|^{2}=10^{22}-10^{25}cm^{-3}$ is the electron's probability
density on a nucleus. The theory presented here is valid in the
opposite limit, $B\gg 100$ G.

This paper is organized in two parts: General theory and applications.
In the first part (sec. II) we describe our theory of spectral
diffusion due to a dipolar coupled spin system.  This general theory
can be easily applied to other spin resonance experiments, such as
three pulse echoes. In the next part (sec. III) we give numerical
results for three particular cases, and discuss their implications for
the current experimental effort in semiconductor spin quantum
computation.  We conclude in sec. IV with a summary and some general
comments.

\section{General Theory}

\subsection{Stochastic theory for the nuclear bath}

Our problem is to describe the dynamics of a carrier spin ${\mathbf
S}$ coupled to a lattice of nuclear spins ${\mathbf I_{n}}$. The total
Hamiltonian can be separated into three parts: ${\cal H}={\cal
H}_{S}+{\cal H}_{SI}+{\cal H}_{I}$,
\begin{equation}
{\cal H}_{S}=\gamma_{S}B S_{z},
\label{hspin}
\end{equation}
\begin{equation}
{\cal H}_{SI}=\sum_{n}A_{n}I_{nz}S_{z},
\end{equation}
\begin{eqnarray}
{\cal H}_{I}&=&-\gamma_{I}B\sum_{n}I_{nz}-4
\sum_{n<m}b_{nm}I_{nz}I_{mz} \nonumber \\ &&+\sum_{n<m}b_{nm}(
I_{n+}I_{m-}+I_{n-}I_{m+}),
\end{eqnarray}
\begin{eqnarray}
A_{n}&=& \gamma_{S}\gamma_{I}\hbar \biggl\{
\frac{8\pi}{3}|\Psi({\mathbf R_{n}})|^{2}\nonumber\\  && - \int d^{3}r
|\Psi({\mathbf r})|^{2} \frac { |{\mathbf r}-{\mathbf R}_{n}|^{2}
-3[({\mathbf r}-{\mathbf R}_{n}) \cdot \hat{{\mathbf z}}]^{2} }
{|{\mathbf r}-{\mathbf R}_{n}|^{5}} \biggr\}
\label{anintegral}\\
&\approx& \gamma_{S}\gamma_{I}\hbar \biggl\{
\frac{8\pi}{3}|\Psi({\mathbf R_{n}})|^{2}\nonumber\\  &&-
\frac{1-3\cos^{2}{\theta_{n}}}{|{\mathbf R}_{n}|^{3}} \theta(|{\mathbf
R}_{n}|-r_{0})\biggl\} ,\label{an}
\end{eqnarray}
\begin{equation}
b_{nm}=-\frac{1}{4}\gamma_{I}^{2}\hbar\frac{1-3\cos^{2}\theta_{nm}}{R_{nm}^{3}}.
\label{bnm}
\end{equation}
Here the Hamiltonians are divided by $\hbar$ to simplify the notation;
$\gamma_{S}$ and $\gamma_{I}$ are gyromagnetic ratios, $A_{n}$ the
coupling with a nucleus located at position ${\mathbf R_{n}}$,
${\mathbf R_{nm}}$ the relative vector between two nuclei and
$\theta_{nm}$ the angle between this vector and the $B$ field
direction.  The electron-nucleus coupling $A_{n}$ includes a Hyperfine
term and a residual dipolar interaction. The Hyperfine term comes from
the singularity of the integral [Eq. (\ref{anintegral})], which is
removed by integrating over the angular coordinates first. Here we
will assume this dipolar term is only effective for  $|{\mathbf
R}_{n}|>r_{0}$, which is a proper length scale for the electron's
wavefunction ($\theta$ is the step function, while $\theta_{n}$ the
angle between ${\mathbf R}_{n}$ and the $B$ field).  The nuclear spins
are in constant turmoil due to their mutual dipolar interaction. To
see how this affects the spin $S$ we approximate each nuclear spin
operator $2I_{nz}$ by a classical random variable $\sigma'_{n}(t)=\pm
1$, which is valid only if the nuclei have spin $1/2$ (this
description is still accurate for spin $I>1/2$, as long as the nuclei
are in the slow SD regime, see below). Hence the Zeeman frequency,
$\omega_{z}$, of the electron spin becomes
\begin{equation}
\omega_{z}(t)=\gamma_{S}B + \frac{1}{2}\sum_{n}A_{n}\sigma'_{n}(t).
\label{zeemanfreq}
\end{equation}
On the other hand, the evolution of the nuclei is strongly affected by
the field produced by the central spin $S$. This effect is treated by
assuming the nuclei evolve according to the effective Hamiltonian
\begin{equation}
{\cal H}'_{I}={\cal H}_{I}+\frac{1}{2}\sum_{n}A_{n}I_{nz},
\label{hiprime}
\end{equation}
which conserves total spin in the $z$ direction. Therefore when any
$I_{nz}$ flips, a corresponding $I_{mz}$ must flop in the opposite
direction. These flip-flop events show that we can not treat the
random variables $\sigma'_{n}$ as independent of each other. Rather,
we will treat pairs of spins as independent random variables. Hence
Eq. (\ref{zeemanfreq}) becomes
\begin{equation}
\omega_{z}(t)=\sum_{n<m}\Delta_{nm}\sigma_{nm}(t)+ const ,
\label{omega_flipflop}
\end{equation}
with $\Delta_{nm}=|A_{n}-A_{m}|/2$, and $\sigma_{nm}=\pm 1$ random
variables uncorrelated with each other. We further make the Markovian
assumption that the probability that $\sigma_{nm}$ changes sign during
a time interval $t$ is given by $t/T_{nm}$, independent of past values
of $\sigma_{nm}$ (this Markovian approximation is reasonable in the
absence of any contrary evidence about the stochastic fluctuations of
the nuclear spins). Hence the number of flip-flops $N(t)$ is a Poisson
random variable with parameter $t/T_{nm}$, and we may write
\begin{equation}
\sigma_{nm}(t)=\sigma_{nm}(0)(-1)^{N(t)},
\label{sigmanm}
\end{equation}
with $N(t)$ having the distribution
\begin{equation}
P(N(t)=k)=\frac{1}{k!}\left(\frac{t}{T_{nm}}\right)^{k}\exp{\left(-\frac{t}{T_{nm}}\right)}.
\end{equation}
In the next section we show how to calculate the flip-flop rate
$T_{nm}^{-1}$.

We now proceed to the derivation of the spin echo decay.  The complex
in-plane magnetization
\begin{equation}
v(t)=\langle S_{x} \rangle + i \langle S_{y}\rangle,
\label{vt}
\end{equation}
can be calculated for any spin echo sequence by taking the
average\cite{klauder62,herzog56}
\begin{equation}
v(t)=\left<\exp{\left( i\int_{0}^{t}s(t')\omega_{z}(t')dt'
\right)}\right>.
\label{echoavg}
\end{equation}
For the $\pi/2-\pi$ echo considered here the echo function $s(t)=1$
for $0\leq t<\tau$ and $s(t)=-1$ for $\tau\leq t$. Therefore
\begin{equation}
v(t)=\prod_{n<m}v_{nm}(t),
\label{product}
\end{equation}
\begin{equation}
v_{nm}(t)= \left<\cos\left[
\Delta_{nm}\int_{0}^{t}s(t')(-1)^{N(t')}dt'\right]\right>,
\label{cos}
\end{equation}
where we take an average over $\sigma_{nm}(0)=\pm 1$ with probability
$1/2$ (this average applies to an ensemble of spins; however, we show
in Appendix B that the effect of the echo is precisely to remove this
average, making $v(t)$ exactly equal to a single spin in-plane
magnetization).  In Appendix A we calculate this average, and the
result is\cite{note1}
\begin{eqnarray}
v_{nm}(t)&=&\theta(\tau - t)
v_{nm}^{(F)}(t)+\theta(t-\tau)v_{nm}^{(E)}(t),
\label{vnmgeneral}\\
v_{nm}^{(F)}(t)&=& \exp{\left( -\frac{t}{T_{nm}}\right)}\left[
\frac{1}{R_{nm}T_{nm}}\sinh{(R_{nm}t)}\right. \nonumber\\ && +
\cosh{(R_{nm}t)}\biggr],
\label{vnmf}\\
v_{nm}^{(E)}(t)&=&R_{nm}^{-2}\exp{\left(
-\frac{t}{T_{nm}}\right)}\left\{ \frac{1}{T_{nm}^{2}}\cosh{(R_{nm}t)}
\right.   \nonumber \\ &&+\frac{R_{nm}}{T_{nm}} \sinh{(R_{nm}t)}
\nonumber  \\ &&-\Delta_{nm}^{2}\cosh{[R_{nm}(t-2\tau)]}\biggr\},
\label{vnm}
\end{eqnarray}
with $R_{nm}^{2}=T_{nm}^{-2}-\Delta_{nm}^{2}$.  We distinguish two
limits in the above expressions: For nuclear pairs causing fast
spectral diffusion, consider  $T_{nm}^{-1}\gg \Delta_{nm}$. In that
case we have
\begin{eqnarray}
v_{nm}^{(E)}(t)\approx v_{nm}^{(F)}(t) &\approx&
\exp{\left(-\frac{1}{2}\Delta_{nm}^{2}T_{nm}t\right)}\nonumber\\ &&+
{\cal O}(\Delta_{nm}^{2}T_{nm}^{2}),
\label{fastvnm}
\end{eqnarray}
which shows that fast flip-flopping nuclei do not even form an echo.
Rather, they contribute to a ``motional narrowed''
signal,\cite{slichter_abragam} which decays exponentially in a time
scale that lengthens as $T_{nm}^{-1}$ increases. The idea is that the
echo sequence can not refocus the spins if their Zeeman frequency
changes appreciably over $\tau$. For large $T_{nm}^{-1}$ the nuclear
pair contributes to a homogeneously broadened line, not an echo
peak.\cite{zhidomirov69}  The slow SD limit $T_{nm}^{-1}\ll
\Delta_{nm}$ implies
\begin{eqnarray}
v_{nm}^{(E)}(t)&=&\exp{\left( -\frac{t}{T_{nm}}\right)} \biggl\{
\cos{[\Delta_{nm}(t-2\tau)]}\nonumber \\ &&+
\frac{1}{\Delta_{nm}T_{nm}}\sin{(\Delta_{nm}t)}\biggr\} \nonumber \\
&&+ {\cal O}(\Delta_{nm}^{-2}T_{nm}^{-2}).
\end{eqnarray}
After performing the product over these slow nuclear pairs,
Eq. (\ref{product}) will have a peak at $t \approx 2\tau$. This is
because if $t\neq 2\tau$, the argument of the product will be zero for
some pair $n,m$ making the whole product vanish. Assuming the echo
peak occurs exactly at $2\tau$, we get
\begin{equation}
v_{nm}^{(E)}(2\tau)\approx \exp{\left\{ -\frac{1}{T_{nm}}\left[ 2\tau
- \frac{\sin{(2\Delta_{nm}\tau)}}{\Delta_{nm}} \right]\right\}}.
\label{slowvnm}
\end{equation}
If in addition to this slow spectral diffusion regime, we look at the
limit $\Delta_{nm}\tau\ll 1$, the result is
\begin{equation}
v_{nm}^{(E)}(2\tau)\approx \exp{\left[
-\frac{1}{6}\frac{\Delta_{nm}^{2}}{T_{nm}}(2\tau)^{3}\right]},
\label{slowvnm2}
\end{equation}
which is very similar to an expression derived  previously by
us\cite{desousa02} using Gaussian spectral diffusion as a starting
point. Apart from a factor of 2 these expressions differ by  a term
which takes into account the small broadening due to spin-flip
processes. Since this term was introduced heuristically and only
changes numerical values by a negligible amount we will not include it
here.

Notice that nothing should be concluded about the qualitative decay of
$v(2\tau)$ before performing the product over pairs in
Eq. (\ref{product}). For example, in the particular case of  SD caused
by dilute paramagnetic impurities  ($\Delta_{nm}\sim r^{-3}$ with $r$
being the impurity-electron distance) it can be shown that
$v(2\tau)\sim \exp{(-a\tau-b\tau^{2})}$\cite{zhidomirov69} after
calculating Eq. (\ref{product}) and taking a spatial average.  Indeed
exactly this behavior was seen in the electron spin echo decay of Si:P
when the P concentration was high enough such that spectral diffusion
due to nearby non-resonant electrons was dominant.\cite{chiba72} Here
we deal with even more complicated expressions for $\Delta_{nm}$ [our
Eqs. (\ref{ansip}), (\ref{ansip31}),  and (\ref{anGaAs})]. Moreover it
is often the case that $T_{nm}^{-1}\sim \Delta_{nm}$ and use of the
limits (\ref{fastvnm}), (\ref{slowvnm}) becomes unjustified. Hence in
our calculations below we perform the product (\ref{product}) using
the exact expressions (\ref{vnm}).

Finally, it is easy to calculate the correlation function [see
Eq. (\ref{avgxi})]
\begin{eqnarray}
\chi_{c}(t)&=&\lim_{t'\rightarrow 0}\frac{\langle
\omega(t)\omega(t')\rangle - \langle \omega(t)\rangle \langle
\omega(t')\rangle} {\langle\omega^{2}(t')\rangle-\langle
\omega(t')\rangle^{2}}
\label{def_correlator}\\ &=&\sum_{n<m}P_{nm}\exp{\left(
-\frac{2t}{T_{nm}}\right)},
\label{correlator}\\
P_{nm}&=&T_{nm}^{-1}\Delta_{nm}^{2}\biggl/
\sum_{i<j}T_{ij}^{-1}\Delta_{ij}^{2}.
\end{eqnarray}
The averages in Eq. (\ref{def_correlator}) assume
$\sigma_{nm}(t=0)=+1$ for all pairs $n,m$.  Then
$\chi_{c}(t_{c})=e^{-1}$ defines the correlation time for the effect
of the nuclear bath on the spin S.   While the two parameter Gaussian
SD theory has a simple exponential correlator
$\chi_{c}=\exp{(-t/T_{1})}$, Eq. (\ref{correlator}) in general does
not decay exponentially. This correlator makes clear the difference
between our many parameter ($T_{nm}$, $\Delta_{nm}$) stochastic theory
and these simpler theories.  If $t \ll t_{c}$, we will have
$\chi_{c}(t)\sim 1$, each $\sigma_{nm}(t)$  performing less than one
flip-flop on average.  Below we show numerically that this regime
leads to $v_{E}(2\tau)\sim \exp{(-\tau^{3})}$. Therefore, when our
calculated  $T_{M}\ll t_{c}$, we predict a Gaussian SD decay for the
echo. However, when $T_{M}\gg t_{c}$, $\chi_{c}(T_{M})\ll 1$
indicating that most $\sigma_{nm}$ have already performed many
flip-flops. In this case we obtain $v_{E}(2\tau)\sim\exp{(-\tau)}$,
characterizing a motional narrowing regime since the nuclear bath
dynamics may be considered fast. Therefore the importance of our
correlation  function is to describe the transition to motional
narrowing induced by all nuclear pairs, not just a single one
satisfying $T_{nm}^{-1}\gg \Delta_{nm}$.

It is important to discuss the case of general nuclei with spin
$I>1/2$, where Eq. (\ref{sigmanm}) does not apply. Then $\sigma(t)$
will be a random walk variable with equal probabilities of moving
right or left,
\begin{equation}
\sigma(t)=2I\Delta, (2I-2)\Delta, \ldots, -2I\Delta .
\label{randomwalk}
\end{equation}
Of course for $I>1/2$ we can not represent $\sigma(t)$ as an
analytical function of the number of steps $N(t)$. However, for small
$t$ (such that $\chi_{c}(t)\approx 1$, or $t T_{nm}^{-1}\ll t
\Delta_{nm}\ll 1$) Eq. (\ref{slowvnm2}) will still hold, since
$\sigma(t)$ will take at most one step (of length $2\Delta$). We will
use this fact when considering GaAs nuclei, which have $I=3/2$ (see
below).
 
\subsection{Calculating the nuclear flip-flop rates}

We now show how to calculate the rate $T_{nm}^{-1}$ appearing in
Eq. (\ref{vnm}), for dipolar coupled spin $I$ nuclei.  As noted above,
we will assume the nuclear dynamics to be decoupled from the central
spin $S$ through the effective Hamiltonian ${\cal H}_{I}^{'}$,
Eq. (\ref{hiprime}). Furthermore, we will separate ${\cal H}_{I}^{'}$
in a secular part ${\cal H}_{0}$ and flip-flop terms $F_{nm}(t)$ which
contain an harmonic time dependence:
\begin{equation} 
{\cal H}_{I}^{'}={\cal H}_{0}+\sum_{n<m} F_{nm}(t),
\end{equation}
\begin{equation} 
{\cal H}_{0}=-\sum_{n}\left(\gamma_{I}B - \frac{1}{2}A_{n}\right)
I_{nz} -4 \sum_{n<m}b_{nm}I_{nz}I_{mz},
\label{h0} 
\end{equation} 
\begin{equation} 
F_{nm}(t) = b_{nm} \left(I_{n+}I_{m-}+I_{n-}I_{m+}\right)\cos{(\omega
t)},
\label{fnm}
\end{equation} 
where we have introduced a fictitious frequency variable $\omega$  for
theoretical convenience -- eventually we are interested in the $\omega
= 0$ limit.  We shall see that it is useful to introduce the frequency
$\omega$: The rate $T_{nm}^{-1}$ will be obtained by taking the
$\omega\rightarrow 0$ limit.  Suppose we have $N$ nuclear spins
${\mathbf I_{n}}$.  We will label the eigenstates of ${\cal H}_{0}$ by
the index $a=0,1,\ldots,2^{N}$: ${\cal H}_{0}|a\rangle =
E_{a}|a\rangle$. Obviously we simply have $|a\rangle =
|m_{1}m_{2}\cdots m_{N}\rangle$ with $m_{i}=\pm 1/2$. The transition
rate between two of these states induced by $F_{nm}(t)$ is then given
by Fermi's golden rule to be
\begin{eqnarray}
{\cal W}_{a,b}^{n,m}&=&\frac{\pi}{2}b_{nm}^{2}\left| \langle a |
F_{nm} |b\rangle\right|^{2}\left[  \delta\left( E_{a}-E_{b}-\omega
\right)  \right.\nonumber \\
&&+\left.\delta\left(E_{a}-E_{b}+\omega\right)\right],
\label{abrate}
\end{eqnarray} 
with $F_{nm}=\left(I_{n+}I_{m-}+I_{n-}I_{m+}\right)$.   The central
spin phase changes by $\Delta_{nm}=\left|A_{n}-A_{m}\right|/2$ during
one of these events.  Moreover, we will assume the nuclei in thermal
equilibrium, each state $|a\rangle$ populated with a Boltzmann
probability
\begin{equation}
p(a)=\exp{\left(-\frac{E_{a}}{k_{B}T}\right)}\biggl/\sum_{b}
\exp{\left(-\frac{E_{b}}{k_{B}T}\right)}.
\end{equation}
Then the flip-flop rate for a pair $nm$ becomes
\begin{equation} 
T_{nm}^{-1}(\omega)=\sum_{a,b}p(a){\cal W}_{a,b}^{n,m}.
\label{tnmomega} 
\end{equation} 
At zero temperature $p(a)$ will be non-zero only for the ground
state. Furthermore, if $B\gg b_{nm}/\gamma_{I}I\sim 0.1$ G this ground
state will simply be $|0\rangle=|I,I,\cdots,I\rangle$, and
$T_{nm}^{-1}=0$ for all pairs $nm$, leading to vanishing zero
temperature decoherence.   However, nuclear Zeeman energies are only
$\sim 1$ mK/Tesla, much lower than typical dilution refrigerator
temperatures ($\gtrsim 10$ mK). Hence a reasonable approximation is to
assume $k_{B}T\gg \hbar \gamma_{I}B$, and $p(a)=(2I+1)^{-N}$ for all
states $|a\rangle$. This in fact makes Eq. (\ref{tnmomega}) much
easier to calculate, particularly since we will employ a technique
similar to the method of moments due to Van
Vleck.\cite{slichter_abragam,vanvleck48} Therefore we assume $p(a)$ to
be the same for all states $a$, and rewrite Eq. (\ref{tnmomega}) in
the form
\begin{equation}
T_{nm}^{-1}(\omega)=2\pi b_{nm}^{2}\rho_{nm}(\omega),
\label{tnmrho}
\end{equation}
with $\rho_{nm}(\omega)$ playing the role of a density of states given
by
\begin{equation}
\rho_{nm}(\omega)=\frac{1}{2(2I+1)^{N}}\sum_{a,b}\left| \langle a |
F_{nm}|b\rangle\right|^{2}\delta\left( E_{a}-E_{b}-\omega \right).
\label{rhonm}
\end{equation}
The invariance of the trace operation allows us to calculate any
moment of $\rho_{nm}(\omega)$ exactly. We define the n-th moment as
\begin{equation}
\langle \omega^{n}\rangle =
\int_{0}^{\infty}\omega^{n}\rho_{nm}(\omega)d\omega \biggl/
\int_{0}^{\infty}\rho_{nm}(\omega)d\omega.
\end{equation}
As an example of how to calculate these moments, consider the
normalization constant
\begin{eqnarray}
{\cal A}(I)&=&\int_{0}^{\infty}\rho_{nm}(\omega)d\omega \nonumber\\
&=&\frac{1}{4(2I+1)^{N}} \sum_{a,b}\langle a | F_{nm}|b\rangle \langle
b | F_{nm}|a\rangle \nonumber \\
&=&\frac{1}{4(2I+1)^{N}}\Tr\left\{F_{nm}^{2}\right\}\nonumber\\
&=&\frac{2}{15}\frac{I(I+1)}{2I+1}[2I(I+1)+1].
\end{eqnarray}
It is straightforward to prove the following relations:
\begin{eqnarray}
\langle \omega \rangle =\overline{\omega} &=& 2{\cal C}(I)\Tr\left\{
\left[ {\cal H}_{0},F_{nm}\right]I_{n+}I_{m-}\right\} \nonumber\\
&=&\frac{1}{2}\left| A_{n}-A_{m}\right|,
\label{avgomega}
\end{eqnarray}
\begin{equation}
{\cal C}(I)=\frac{15}{8}\frac{(2I+1)^{-N+1}}{I(I+1)[2I(I+1)+1]},
\end{equation}
\begin{eqnarray}
\langle\omega^{2}\rangle &=& -{\cal C}(I) \Tr\left\{ \left[{\cal
H}_{0},F_{nm}\right]^{2}\right\},\\  \langle \left( \omega -
\overline{\omega} \right)^{2}\rangle  &=&
\frac{16}{3}I(I+1)\sum_{i\neq n,m}\left(b_{ni}-b_{mi}\right)^{2},
\label{seconmom}
\end{eqnarray}
\begin{equation}
\langle \omega^{4}\rangle = {\cal C}(I)\Tr\left\{ \left[{\cal
H}_{0},\left[{\cal H}_{0},F_{nm}\right]\right]^{2}\right\},
\end{equation}
\begin{eqnarray}
\langle \left( \omega - \overline{\omega}\right)^{4}\rangle &=&
\frac{2^{8}}{3}I(I+1)\biggl\{ -\frac{1}{5}[2I(I+1)+1]\nonumber\\
&&\times\sum_{i\neq n,m}\left(b_{ni}-b_{mi}\right)^{4}\nonumber\\ &&+
I(I+1)\left[\sum_{i\neq
n,m}\left(b_{ni}-b_{mi}\right)^{2}\right]^{2}\biggr\}.
\label{fourthmom}
\end{eqnarray}
Using these expressions we can discuss various possibilities for  the
shape of $\rho_{nm}(\omega)$, which is clearly an even function of
$\omega$ with peaks at $\pm |A_{n}-A_{m}|/2$. The most common choice
in lineshape theory is to assume each peak to be a Gaussian or a
Lorentzian with cut-off at the wings.\cite{slichter_abragam}  To
decide between these, we define a dimensionless parameter
\begin{eqnarray}
\xi_{nm}&=&\frac{\langle \left( \omega -
\overline{\omega}\right)^{4}\rangle}{3\langle \left( \omega -
\overline{\omega}\right)^{2}\rangle^{2}} \nonumber \\ &=&
1+\left(\frac{1-3f}{3f}\right)\frac{\sum_{i\neq
n,m}\left(b_{ni}-b_{mi}\right)^{4}}{\left[\sum_{i\neq
n,m}\left(b_{ni}-b_{mi}\right)^{2}\right]^{2}}.
\label{ratiomoments}
\end{eqnarray}
Here we have introduced the occupation probability $f$ for a nucleus
at site $i$ (sums involving one site and two sites are proportional to
$f$ and $f^{2}$ respectively).  For a Gaussian function this ratio
would be exactly 1. For a Lorentzian, it would be equal to $\sim
\alpha_{nm}/\delta_{nm} \gg 1$, where $\alpha_{nm}$ is the cut-off and
$\delta_{nm}$ the half-width at half maximum. Clearly $f\ll 1$ leads
to Lorentzian rates.  Here we will adopt the following approximation:
Whenever $\xi_{nm} \leq 10$ we will approximate $\rho_{nm}(\omega)$ by
a Gaussian; if $\xi_{nm} > 10$ we will use the Lorentzian fit. The
Gaussian approximation leads to
\begin{eqnarray}
\rho_{nm}(\omega)&=&{\cal A}(I)\frac{1}{\sqrt{2\pi}\kappa_{nm}}\left\{
\exp{\left[ -\frac{(\omega
-\overline{\omega})^{2}}{2\kappa_{nm}^{2}}\right]} \right.\nonumber \\
&&\left.+\exp{\left[ -\frac{(\omega
+\overline{\omega})^{2}}{2\kappa_{nm}^{2}}\right]} \right\},
\label{rhonmgauss}
\end{eqnarray}
with
\begin{equation}
\kappa_{nm}=\sqrt{\langle(\omega-\overline{\omega})^{2}\rangle}.
\end{equation}
The desired flip-flop rate is then obtained by setting $\omega = 0$ in
Eqs. (\ref{rhonmgauss}) and (\ref{tnmrho}),
\begin{equation}
T_{nm}^{-1}(\xi_{nm}\leq 10)=2\sqrt{2\pi}{\cal A}(I)\frac{b_{nm}^{2}}
{\kappa_{nm}}\exp{\left(
-\frac{|A_{n}-A_{m}|^{2}}{8\kappa_{nm}^{2}}\right)}.
\label{tnmgaussfinal}
\end{equation}
For $\xi_{nm}>10$ we assume the Lorentzian form
\begin{eqnarray}
\rho_{nm}(\omega)&=&{\cal A}(I)\frac{\delta_{nm}}{\pi}\nonumber \\
&&\times\left[\frac{\theta(\omega -
\overline{\omega}+\alpha_{nm})-\theta(\omega
-\overline{\omega}-\alpha_{nm})}{\delta_{nm}^{2}+(\omega -
\overline{\omega})^{2}}\right.\nonumber\\ &&+
\left. \frac{\theta(\omega +
\overline{\omega}+\alpha_{nm})-\theta(\omega
+\overline{\omega}-\alpha_{nm})}{\delta_{nm}^{2}+(\omega +
\overline{\omega})^{2}}  \right],
\label{rhonmlor}
\end{eqnarray}
with $\theta$ being the step function.   It is important to note that
this Lorentzian must have a cut-off so that its second and higher
moments do not diverge. The parameters $\alpha_{nm}$ and $\delta_{nm}$
can be easily related to Eqs. (\ref{seconmom}) and (\ref{fourthmom}):
\begin{eqnarray}
\delta_{nm} &\approx& \frac{\pi}{2\sqrt{3}}\sqrt{\frac{\langle \left(
\omega - \overline{\omega}\right)^{2}\rangle^{3}}{\langle \left(
\omega - \overline{\omega}\right)^{4}\rangle}},\label{delta} \\
\alpha_{nm} &\approx& \sqrt{\frac{2\langle \left( \omega -
\overline{\omega}\right)^{4}\rangle}{\langle \left( \omega -
\overline{\omega}\right)^{2}\rangle}}.
\label{alpha}
\end{eqnarray}
The Lorentzian flip-flop rate is
\begin{equation}
T_{nm}^{-1}(\xi_{nm}>10)=4{\cal A}(I)\frac{b_{nm}^{2}}{\delta_{nm}}
\frac{1-\theta(|A_{n}-A_{m}|- 2\alpha_{nm})}
{1+\left[(A_{n}-A_{m})/(2\delta_{nm})\right]^{2}}.
\label{tnmfinal}
\end{equation}
Therefore $T_{nm}^{-1}$ will be suppressed unless
$|A_{n}-A_{m}|\lesssim \kappa_{nm},\delta_{nm}$. This effect stems
from energy conservation: For a flip-flop to happen, an amount
$|A_{n}-A_{m}|/2$ of energy must be absorbed by the dipolar term in
${\cal H}_{0}$. To see this consider two states with $I=1/2$, \newline
$|a\rangle=|+1/2,-1/2,m_{3},\ldots,m_{N}\rangle$, \newline
$|b\rangle=|-1/2,+1/2,m_{3},\ldots,m_{N}\rangle$. Their energy
difference is
\begin{equation}
E_{a}-E_{b}=\frac{1}{2}\left( A_{1}-A_{2}\right)-4\sum_{i\neq
1,2}\left( b_{1i}-b_{2i}\right)m_{i}.
\end{equation}
If we calculate the average square of this expression using $\langle
m_{i}^{2}\rangle = 1/4$, $\langle m_{i}\rangle = 0$ we get
\begin{equation}
\langle \left(E_{a}-E_{b}\right)^{2} \rangle = \left(
\frac{A_{1}-A_{2}}{2}\right)^{2}+4\sum_{i\neq 1,2}\left(
b_{1i}-b_{2i}\right)^{2},
\end{equation}
showing that an adjustment of the spin system can supply energies up
to $\sim \kappa_{nm}$ to compensate for the central spin field.  We
believe that the above expressions for $T_{nm}^{-1}$  are considerably
more precise than the phenomenological rates used in the  earlier
literature without any
derivations.\cite{desousa02,chiba72,bloembergen59}


\section{Applications}

\subsection{Electron spin of a phosphorus donor in silicon}

We now turn to applications of our theory for systems of interest to
quantum computation in spin qubit-based semiconductor architectures.
We start by considering the electron spin of a shallow donor in
silicon. In natural samples, 95.33\% of silicon atoms have no nuclear
magnetic moment:\cite{handbook} Those are the $^{28}$Si
isotopes. Spectral diffusion is then caused by the remaining fraction
$f_{nat}=0.0467$ of $^{29}$Si isotopes which are spin $1/2$ nuclei
with gyromagnetic ratio $\gamma_{I}^{Si}=5.31\times 10^{3}$ (s
G)$^{-1}$. These nuclei will produce a hyperfine field on the electron
donor impurity given by Eq. (\ref{an}), which is proportional to the
electron's probability density at the nuclear site ${\mathbf R_{n}}$,
$|\Psi({\mathbf R_{n}})|^{2}$. For this state we assume the
Kohn-Luttinger wave function\cite{feher59a,kohn57}
\begin{eqnarray}
\Psi({\mathbf r}) &=& \frac{1}{\sqrt{6}}\sum_{j=1}^{6}F_{j}({\mathbf
r})u_{j}({\mathbf r})e^{i {\mathbf k_{j}\cdot r}},
\label{kohnluttinger}
\\ {\mathbf k_{j}}&=& 0.85 \frac{2\pi}{a_{Si}}\hat{k}_{j} , \;
\hat{k}_{j} \in \left\{
\hat{x},-\hat{x},\hat{y},-\hat{y},\hat{z},-\hat{z}\right\},\\
F_{1,2}({\mathbf r}) &=& \frac{\exp{\left[
-\sqrt{\frac{x^{2}}{(nb)^{2}}+\frac{y^{2}+z^{2}}{(na)^{2}}}\right]}}
{\sqrt{\pi(na)^{2}(nb)}}\label{f12},
\end{eqnarray}
with the appropriate corresponding envelope functions $F_{j}$
[Eq. (\ref{f12})] with anisotropies in the $y$ and $z$
directions. Here $n=(0.029eV/E_{i})^{1/2}$ with $E_{i}$ being the
ionization energy of the impurity ($E_{i}=0.044$ eV for the phosphorus
impurity, hence $n=0.81$ in our case), $a_{Si}=5.43$ \AA\ the lattice
parameter for Si, $a=25.09$ \AA\ and $b=14.43$ \AA\ characteristic
lengths for Si hydrogenic impurities.\cite{feher59a} Moreover, we will
use experimentally measured values for the charge density on each Si
lattice site,\cite{kohn57,shulman56}
\begin{equation}
|u_{j}({\mathbf R_{n}})|^{2}=\eta\approx 186.
\end{equation}
Hence the hyperfine interaction is given by
\begin{eqnarray}
A_{n}&=&\frac{16\pi}{9}\gamma_{S}^{Si} \gamma_{I}^{Si}\hbar\eta\left[
F_{1}({\mathbf R_{n}})\cos{(k_{0}X_{n})} \right.\nonumber\\
&&\left.+F_{3}({\mathbf R_{n}})\cos{(k_{0}Y_{n})} +F_{5}({\mathbf
R_{n}})\cos{(k_{0}Z_{n})}\right]^{2}\nonumber \\ &&-
\gamma_{S}^{Si}\gamma_{I}^{Si}\hbar
\frac{1-3\cos^{2}{\theta_{n}}}{|{\mathbf R}_{n}|^{3}} \theta(|{\mathbf
R}_{n}|-na) ,\label{ansip}
\end{eqnarray}
with $k_{0}=(0.85)2\pi/a_{Si}$, and $\gamma_{S}^{Si}= 1.76\times
10^{7}$ (s G)$^{-1}$ the gyromagnetic ratio for the electron donor. It
is instructive to check the experimental validity of Eq. (\ref{ansip})
by calculating the inhomogeneous line-width ($\sim
1/\gamma_{S}^{Si}T_{2}^{*}$). A simple statistical theory applied to
Eq. (\ref{zeemanfreq}) leads to\cite{kohn57}
\begin{equation}
\langle \left( \omega/\gamma_{S}^{Si}-B\right)^{2}\rangle=
\frac{f}{(2\gamma_{S}^{Si})^{2}}\sum_{{\mathbf R_{n}}\neq {\mathbf
0}}A_{n}^{2}.
\label{T2star}
\end{equation}
For $f=0.0467$ our calculated root mean square line-width is equal to
$0.89$ G. On the other hand, an ESR scan leads to $2.5$
G$/2\sqrt{2\ln{2}}=1.06$ G.\cite{feher59a} Therefore our model is able
to explain 84\% of the experimental hyperfine line-width  [the
residual dipolar term in Eq. (\ref{ansip}) only contributes $\sim
0.1\%$ to this line-width].

Before discussing our spin echo decay results we should mention how
the $^{29}$Si fraction $f$ enters our calculations. For example,
Eq. (\ref{product}) becomes
\begin{equation}
v(2\tau)=\prod_{n<m}\left[v_{nm}(2\tau)\right]^{f^{2}},
\label{productf2}
\end{equation}
since the probability of a pair $n,m$ to be $^{29}$Si is $f^{2}$.
Also single and double sums in Eqs. (\ref{seconmom}) and
(\ref{fourthmom}) are proportional to $f$ and $f^{2}$ (these sums are
calculated assuming the pair $n,m$ is already occupied; hence the
question is whether or not site $i\neq n,m$ contains a $^{29}$Si).  To
perform our numerical calculations we take the natural logarithm of
Eq. (\ref{productf2}). Then we take advantage of the fact that
$b_{nm}\propto R_{nm}^{-3}$ [Eq. (\ref{bnm})] decays fast as a
function of the inter-nuclear distance $R_{nm}$. Therefore we achieve
faster convergence by summing over lattice sites $n$ together with
some of its close neighbors $m$. Both the range of sites $n$ and the
number of neighbors included in the sum were increased systematically
to ensure proper convergence. From the convergence check we concluded
that summing within eight characteristic lengths of the wave function
($\sim 40 a_{Si}$ for the Si:P case) plus including up to six nearest
neighbor shells  gave excellent convergence.

\begin{figure}
\includegraphics[width=3in]{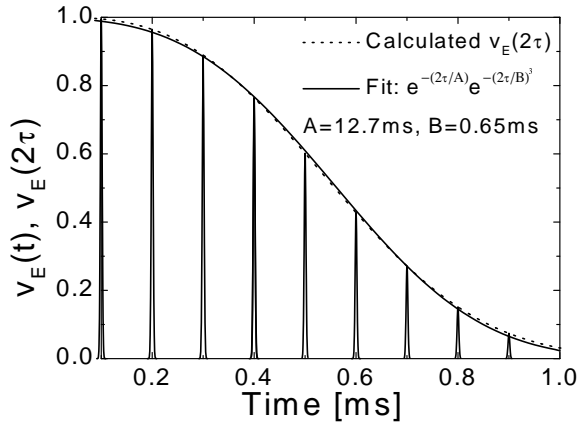}
\caption{ Calculated echo peaks $v_{E}(t)$ and echo envelope
$v_{E}(2\tau)$  as a function of time for a single donor electron spin
in Si:P. We assumed the Si lattice had natural abundance of $^{29}$Si
($f=0.0467$) and $B \parallel [111]$. We also show a fit for our echo
envelope that compares well with the experimental fit of
Ref.~\onlinecite{chiba72} [see Eq. (\ref{chiba_exp})].
\label{figone}}
\end{figure}

Fig. 1 shows the echo signal [Eq. (\ref{vnm})] as a function of time
for natural  Si and $B\parallel [111]$. Clearly the echoes occur at
$t=2\tau$. The echo   envelope, defined by the maximum of each peak,
should be compared with the empirical experimental fit of Chiba and
Hirai,\cite{chiba72}
\begin{equation}
v_{Exp}(2\tau)=\exp{\left[-\left(\frac{2\tau}{0.6
ms}\right)-\left(\frac{2\tau}{0.4 ms}\right)^{3}\right]}.
\label{chiba_exp}
\end{equation}
Clearly our theory is able to explain quite successfully the
$\exp{(-\tau^{3})}$ decay, but our exponential tail is twenty times
smaller than the measured value. Therefore our theory suggests this
extra exponential decay is coming from other decoherence mechanisms,
perhaps related to imperfections of the ESR pulses [Recent
experimental data\cite{tyryshkin03} suggest dipolar scattering between
donor electrons is responsible for the extra exponential decay seen in
Eq. (\ref{chiba_exp})].  Nevertheless, our $T_{M}=0.64$ ms is $\sim 2$
times larger than the measured value of $0.3$ ms, a quite good
agreement in view of former spin relaxation
calculations.\cite{feher59b} Including the residual dipolar term in
$A_{n}$ [see Eq. (\ref{ansip})] changes $T_{M}$ by less than 1\%,
showing that here SD is dominated by flip-flopping nuclei inside the
electron's wave function [if we set the hyperfine term to zero in Eq.
(\ref{ansip}), we get $T_{M}=1.7-1.9$ ms for $0<r_{0}\leq na$].  It is
interesting to note that former SD estimates\cite{mims72,chiba72}
neglected the hyperfine contribution, arguing that nuclei inside the
electron's wave function could not flip-flop due to the large
$\Delta_{nm}$ they would induce. Surprisingly, our calculation shows
that this coupling is very important, although estimates using only
the dipolar term would still lead to reasonable results. The hyperfine
term becomes more and more important as the wave function size
increases.  Fig. 2 shows the behavior of $-\ln{v_{E}(2\tau)}$ for
$B\parallel [111]$ and a few values of $f$. As $2\tau$ is increased,
the echo envelope changes from a ``Gaussian spectral diffusion''
regime\cite{klauder62} [$\sim\tau^{3}$, well described by Eq.
(\ref{slowvnm2})] to $-\ln{v}\sim \tau$ [see Eq. (\ref{fastvnm}) and
(\ref{slowvnm}) for $\Delta_{nm}\tau\gg 1$]. The ``experimental
window'' $0.1\lesssim -\ln{v}\lesssim 3$ falls within this crossover
for many values of $f$, including natural isotopic abundance. The
exponential behavior obtained for long $2\tau$ can be understood by
looking at the correlation function [Eq. (\ref{correlator})] which
also appears in Fig. 2 (right scale). Notice that this correlation
function decays smoothly over 3 time decades, but its behavior is
clearly non-exponential (see Fig. 9), making evident the difference
between our current complete theory and former heuristic ones based on
the two parameter Gaussian conditional
probability.\cite{desousa02,klauder62,herzog56} The small correlation
between Zeeman frequencies before and after the application of the
$\pi$ pulse explains this motional narrowing behavior for $v(2\tau)$
(see discussion in the end of section II-A). Note that Fig. 2 suggests
it might be more appropriate to fit $-\ln{v(2\tau)}$ to a function
which approaches $\tau$ at long times and $\tau^{3}$ at short time
scales. Such a fit could possibly yield a better description for
experimental data as long as other contributing mechanisms
(electron-electron dipolar scattering\cite{desousa02} for example) are
properly subtracted.

\begin{figure}
\includegraphics[width=3in]{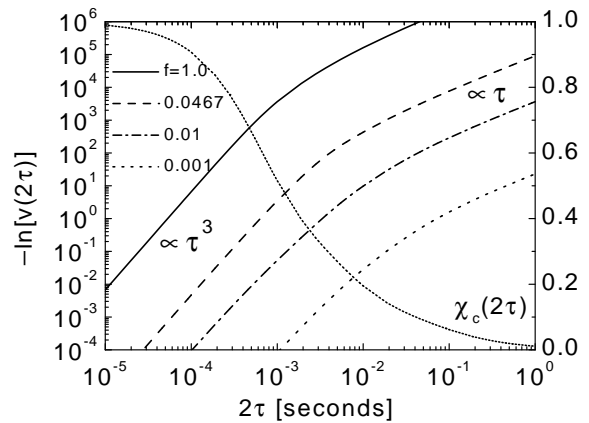}
\caption{ The left scale shows the logarithm of the $\pi/2 - \pi$ echo
envelope as a function of $2\tau$ for electron spins of Si:P. For all
$^{29}$Si isotopic fractions $f$ this function undergoes a crossover
from Gaussian spectral diffusion behavior [$v\sim \exp{(-k\tau^{3})}$]
to motional narrowing [$v\sim \exp{(-k\tau)}$]. The correlation
function,  $\chi_{c}(2\tau)$ is shown on the right scale; motional
narrowing behavior corresponds to small values of $\chi_{c}$. This
correlation function shows non-exponential decay, quite distinct from
former theories\cite{desousa02,klauder62,herzog56}(see also Fig. 9
below). Here we have  $B\parallel [111]$.
\label{figtwo}}
\end{figure}

It is interesting to see if the Si six-fold degeneracy embedded into
the Kohn-Luttinger state [Eq. (\ref{kohnluttinger})] has a strong
effect on $T_{M}$. We obtain $T_{M}\sim 0.3$ ms at natural abundance
using a hydrogenic state with Bohr radius $\sim 20$ \AA. Therefore we
can conclude that the strong oscillations of $A_{n}$ reduce
$T_{nm}^{-1}$ by a small amount, hence increasing $T_{M}$ (as
$\Delta_{nm}$ is larger the energy cost for nuclear flip-flops is
higher). This gives an idea about the effect of the electron's field
on the $^{29}$Si nuclei.  Fig. 3 shows $T_{M}$ as a function of
$f$. Only one purified sample with $f=(0.12\pm 0.08)\%$ was studied
experimentally,\cite{gordon58} leading to $T_{M}=0.52$ ms. Note that
this value is significantly different than what we have in Fig. 3
($\sim 0.1$ s) simply because at such low values of $f$ the
concentration of donors ($4\times 10^{16}$ cm$^{-3}$) lead to strong
electron-electron dipolar scattering (see Fig. 2 of
Ref.~\onlinecite{desousa02}). It would be quite interesting to see
spin echo measurements for different values of $f$, but as the
isotopic purity increases one will have to decrease donor
concentration significantly to ensure that dipolar scattering is not
prevailing (for that purpose $T_{M}$ should be independent of donor
concentration, showing saturation for low P
concentration\cite{chiba72}).

An interesting confirmation of our theory would be to measure $T_{M}$
as a function of the $B$ field tilting angle $\theta$. Fig. 4 shows
this angular dependence in samples with natural isotopic abundance.
$\theta = 0^{\circ}$ means $B\parallel [001]$, while $\theta =
90^{\circ}$ $B\parallel [110]$ (unfortunately
Ref.~\onlinecite{chiba72} measured $T_{M}$ only for $B\parallel
[111]$). Preliminary experimental data shows excellent qualitative
agreement with Fig. 4.\cite{tyryshkin03B} Incidentally, our theory
does not depend on the $B$ field intensity, except for $B\gtrsim 10$
T: Then $k_{B}T\gg \hbar \gamma_{I}B$ does not hold, and also the
magnetic length $l_{B}=(\hbar c/eB)^{1/2}\lesssim 20$ \AA, and B is
already deforming the electron's wave function. The latter effect does
in fact appear in the quantum dot case, since its radius depends
crucially on $l_{B}$ [see Eq. (\ref{fockdarwinradius})].

\begin{figure}
\includegraphics[width=3in]{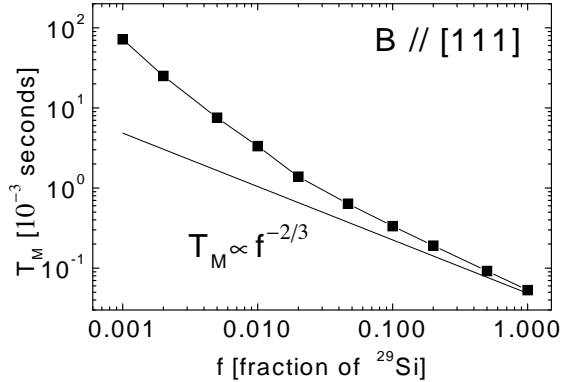}
\caption{Decoherence time ($T_{M}$) of the Si:P electron spin as a
function of $^{29}$Si content $f$. The straight line resembles a
simple theory derived previously\cite{desousa02} where $f$ only
altered the flip-flop pair probability. Our new theory deviates from
this behavior since $f$ also enters the moments [Eqs. (\ref{seconmom})
and (\ref{fourthmom})]. The change in slope seen in $T_{M}$ is
explained by  the transition from Gaussian to Lorentzian  flip-flop
expressions for $T_{nm}^{-1}$ [See Eqs. (\ref{tnmgaussfinal}) and
(\ref{tnmfinal})].
\label{figthree}}
\end{figure}
\begin{figure}
\includegraphics[width=3in]{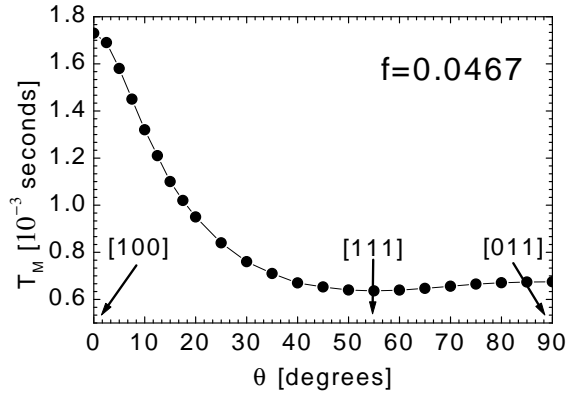}
\caption{ Behavior of Si:P electron spin $T_{M}$ as a function of $B$
field tilting angle with respect to the crystal lattice for samples
with natural abundance of $^{29}$Si. To our knowledge, this dependence
has never been probed experimentally, and would be an interesting test
for the accuracy of our theory.
\label{figfour}}
\end{figure}

\subsection{Nuclear spin of a phosphorus donor in silicon}

$^{31}$P nuclear spins [$\gamma_{S}^{P}=1.08\times 10^{4}$ (s
G)$^{-1}$] are promising qubit candidates when implanted in a Si
matrix.\cite{kane98} Its interaction with the lattice is rather weak,
leading to a measured $T_{1}\sim 5$ hours at $B=0.8$ Tesla and
$T=1.25$ K.\cite{feher59b} Here we calculate the contribution of
$^{29}$Si spectral diffusion on the decoherence time $T_{M}$ of an
isolated $^{31}$P nucleus. There is an important difference between
this calculation and the one related to the Si:P electron spin
above. Even though the flip-flop rate of a pair $n,m$ of $^{29}$Si
($T_{nm}^{-1}$) is determined in exactly the same way as above
[Eqs. (\ref{tnmgaussfinal}) and (\ref{tnmfinal}) with $A_{n}$ given by
Eq. (\ref{ansip})], the corresponding phase change on the $^{31}$P
spin is given by
\begin{eqnarray}
\Delta_{nm}&=& |A_{n}' - A_{m}'|/2,\\ A_{n}'&=&
\gamma_{S}^{P}\gamma_{I}^{Si}\hbar
\frac{1-3\cos^{2}\theta_{n}}{|{\mathbf R_{n}}|^{3}},
\label{ansip31}
\end{eqnarray}
where $A_{n}'$ is the dipolar interaction between a $^{29}$Si nucleus
located at ${\mathbf R_{n}}$ and the $^{31}$P located at the origin
(in principle we should include the effect of the  $^{31}$P dipolar
field on the flip-flop rates $T_{nm}^{-1}$, but we determined it to be
negligible on the $T_{M}$ calculation both in this section and the
preceding one). It is easy to see that this phase change $\Delta_{nm}$
is in most cases $\sim 10^{3}$ smaller than the corresponding one for
the electron. Therefore the motional narrowing condition
$T_{nm}^{-1}\gg \Delta_{nm}$ will be satisfied for many more pairs
than in the last section. Hence we should expect $v\sim
\exp{(-k\tau)}$ for a wide range of parameters [see
Eq. (\ref{fastvnm})].

\begin{figure}
\includegraphics[width=3in]{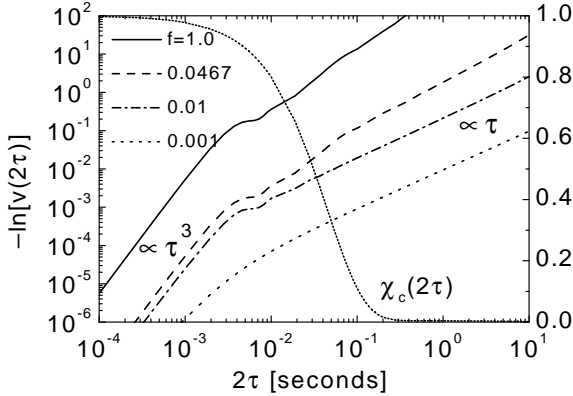}
\caption{Logarithm of $\pi/2 - \pi$ echo envelope as a function of
  echo time $2\tau$ for $^{31}$P nuclear spins in Si:P.  This function
  undergoes an abrupt crossover from Gaussian spectral diffusion
  behavior to motional narrowing, quite different from the smooth
  transition seen on Fig. 2. On the right scale we plotted the
  correlation function $\chi_{c}$, which shows a sharper transition to
  motional narrowing as compared to Fig. 2 (here $\chi_{c}$ changes
  over one time decade, as opposed to three in Fig.~2).
\label{figfive}}
\end{figure}

Fig. 5 shows the calculated shape of the echo envelope as a function
of $2\tau$. The qualitative behavior is similar to Fig. 2 above,
except for the rather abrupt crossover from Gaussian SD to motional
narrowing behavior. For $f\leq 0.0467$ the observed decay is indeed
$v\sim \exp{(-k\tau)}$ as predicted above.  This behavior can be
understood by looking at the correlation function $\chi_{c}$ (right
scale on Fig. 5) which  goes to zero in only one time decade,
evidencing the abrupt appearance of the motional narrowing regime.
Unfortunately there are no experimental data available to verify this
qualitative result. Fig. 6 depicts the dependence of $T_{M}$ as a
function of $f$.

\begin{figure}
\includegraphics[width=3in]{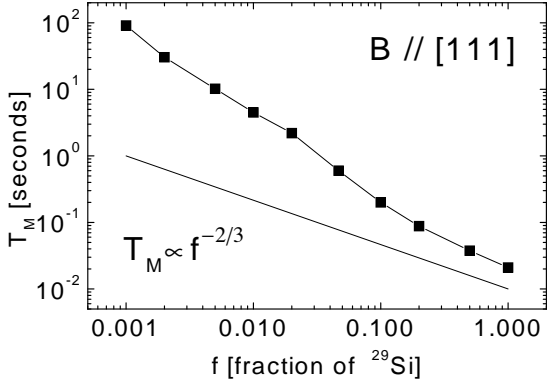}
\caption{Depicts $T_{M}$ versus $^{29}$Si content $f$ for $^{31}$P
nuclear spins. Similar to Fig. 3, the change in slope around $f\sim
0.01$  occurs due to the crossover to Lorentzian $T_{nm}^{-1}$
[Eq. (\ref{tnmfinal})].
\label{figsix}}
\end{figure}

The dependence on the tilting angle $\theta$ (Fig. 7) is significantly
different than the one in Fig. 4. This can be explained by the fact
that while in the former case first nearest-neighbor flip-flop
dominated (about 92\% of $T_{M}^{-1}$), here first nearest neighbor
amounts only to 3\% of the rate, which is dominated by second nearest
neighbors ($\sim 90\%$ of $T_{M}^{-1}$). This is why we see more
oscillations in Fig. 6 than in Fig. 3: Second nearest neighbors are 12
in number, and have a more intricate lattice configuration than the 4
first nearest neighbors. Quite interestingly, in Fig. 7, $T_{M}$ is
maximized when $B$ points in the [110] direction, and displays a sharp
peak in the [111] direction, while Fig. 4 shows $T_{M}$ minimized when
$B$ points in both directions.

\begin{figure}
\includegraphics[width=3in]{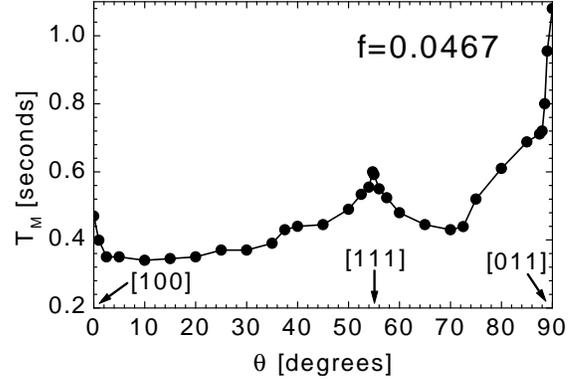}
\caption{Spectral diffusion decay time $T_{M}$ versus $B$ field
tilting angle for $^{31}$P nuclei. This calculation assumes natural
abundance of $^{29}$Si, $f=0.0467$.
\label{figseven}}
\end{figure}

\subsection{Electron spin in a gallium arsenide quantum dot}

Recent interest in the spin properties of a single electron GaAs QD is
motivated by its potential use as a qubit.\cite{loss98,vandersypen02}
One advantage over Si:P is the non-degenerate conduction band, with
the minimum at the $\Gamma$ point.  Therefore the exchange interaction
of a double dot is a smooth function of the inter-dot barrier and
distance, quite different from two Si:P donors, where this exchange
may oscillate dramatically as a function of donor
separation.\cite{koiller02}

The GaAs lattice has a Zincblende structure with 50\% of $^{75}$As,
30.2\% of $^{69}$Ga, and 19.8\% of $^{71}$Ga.\cite{handbook}  These
nuclei have spin $I=3/2$, but here we present calculations using the
spin $1/2$ stochastic theory together with spin $3/2$ flip-flop rates.
Since the correlation function is very close to 1 in the neighborhood
of our calculated $T_{M}$, the electron is well into the slow SD
regime, making this approximation very accurate for our purpose.  The
electron hyperfine interaction with these nuclei will lead to an
inhomogeneous linewidth of about $50$ G for small dots (with
Fock-Darwin radius $\ell \sim 20$ nm; donor impurities in bulk GaAs
have even higher broadening\cite{seck}). If one has an ensemble of
these dots, the decoherence time $T_{M}$ can be measured by using a
$\pi/2-\pi$ pulse. However, ensembles of dots always contain size
distribution. Since $T_{M}$ will be quite sensitive to the radius
$\ell$ (see below), it might be more appropriate to measure
decoherence by applying an ESR field to a single dot and then
measuring the signal using transport experiments, even though such
experiments would only lead to a lower bound on $T_{M}$.\cite{engel01}

The energy eigenstates of a quantum dot in the presence of spin-orbit
coupling are a mixture of spin states up and down.  Therefore if the
electron spin is up, it will flip within  a time $T_{1}$ with the
corresponding emission of a phonon.  However,  for the spin to flip a
virtual transition to an excited orbital state has to happen, since
spins do not couple directly to the phonon strain field. The result is
a strong sensitivity on dot size and applied field $B$,
$T_{1}^{-1}\propto \ell^{8}B^{5}$. For $B\sim 1$ Tesla and $\ell \sim
30$ nm a recent theory leads to $T_{1}\sim 1$ ms, showing that for the
small dots  in a quantum computer architecture spin-flip scattering is
strongly suppressed.\cite{khaetskii01} Recently we showed that the
dominant decoherence mechanism in these small dots  ($\ell \lesssim
50$ nm) is nuclear spectral diffusion.\cite{desousa02} Here, our
detailed calculation of this effect confirms the accuracy of the
simple theory presented earlier,\cite{desousa02} justified by
Eq. (\ref{slowvnm2}) and the fact that the correlation function
[Eq. (\ref{correlator})] is still quite close to $1$ in the
neighborhood of $T_{M}$.  For $|z|\leq z_{0}/2$, the quantum dot wave
function can be simply approximated as
\begin{equation}
\Psi({\mathbf r})=\sqrt{\frac{2}{z_{0}}}\cos{\left(\frac{\pi
}{z_{0}}z\right)} \frac{1}{\sqrt{\pi}\ell(B)}
\exp{\left(-\frac{x^{2}+y^{2}}{2\ell^{2}(B)}\right)},
\label{qdwavefunction}
\end{equation}
and we assume $\Psi=0$ for $z>z_{0}/2$. This state is a reasonable
description for the lowest orbital of a quantum well of thickness
$z_{0}$, with electrostatic lateral parabolic confinement with radius
$\ell_{0}$. The Fock-Darwin radius $\ell(B)$ includes the additional
$B$ field confinement,
\begin{eqnarray}
\ell(B)&=&\frac{l_{B}l_{0}}{\sqrt[4]{l_{B}^{4}+l_{0}^{4}/4}},
\label{fockdarwinradius}\\
l_{B}&=&\sqrt{\frac{\hbar c}{e B}}.
\end{eqnarray}
Then the hyperfine coupling $A_{n}$ for a nucleus located at a
coordinate $(X_{n},Y_{n},Z_{n})$ from the center of the dot becomes
\begin{eqnarray}
A_{n}&=&\frac{16}{3}\frac{\gamma_{S}\gamma_{I}\hbar
a_{GaAs}^{3}}{\ell^{2}(B) z_{0}}d(I)
\cos^{2}{\left(\frac{\pi}{z_{0}}Z_{n}\right)}\nonumber\\
&&\times\exp{\left(-\frac{X_{n}^{2}+Y_{n}^{2}}{\ell^{2}(B)}\right)}
\theta(z_{0}/2 - |Z_{n}|)\\  &&-\gamma_{S}\gamma_{I}\hbar
\frac{1-3\cos^{2}{\theta_{n}}}{|{\mathbf R}_{n}|^{3}}
\theta(X_{n}^{2}+Y_{n}^{2}-\ell^{2}(B)).
\label{anGaAs}
\end{eqnarray}
Here $a_{GaAs}=5.65$ \AA, $\gamma_{S}=- 3.86 \times 10^{6}$ (s
G)$^{-1}$ (assuming $g=-0.44$ independent of $\ell$ for the dot
electron\cite{kiselev98}),  $\gamma_{I}=4.58,\: 8.16,\: 6.42 \times
10^{3}$ (s G)$^{-1}$ and charge densities $d(I)=9.8,\: 5.8,\:
5.8\times 10^{25}$ cm$^{-3}$ for $^{75}$As, $^{71}$Ga, and $^{69}$Ga
respectively.\cite{paget77} The residual dipolar coupling again leads
to a very small correction ($<1\%$ of $T_{M}$).

\begin{figure}
\includegraphics[width=3in]{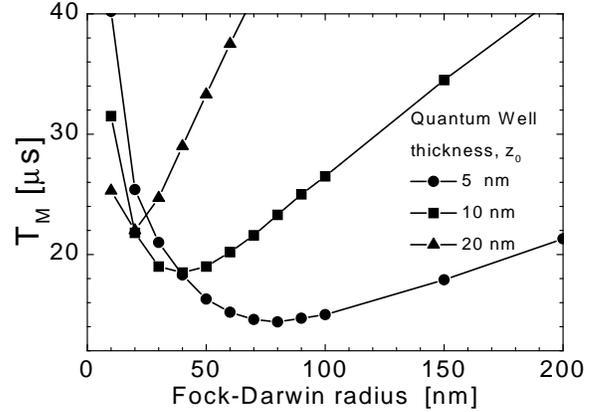}
\caption{ $T_{M}$ versus Fock-Darwin radius $\ell$ for various quantum
well thicknesses, $z_{0}=5,\: 10,\: 20$ nm. Decoherence achieves a
maximum as a function of $\ell$. Here $B\parallel [111]$.
\label{figeight}}
\end{figure}

Fig. 8 shows the behavior of $T_{M}$ as a function of Fock-Darwin
radius $\ell$ for three different quantum well thicknesses $z_{0}$.
For each $z_{0}$, $T_{M}$ displays a minimum as a function of $\ell$.
This arises from two competing effects: If we decrease the wave
function size, fewer nuclei contribute to SD, hence $T_{M}\rightarrow
\infty$ as $\ell \rightarrow 0$. However when we increase $\ell$ the
wave function flattens, making two close pairs $n,m$ produce similar
hyperfine fields. Since it is the difference $|A_{n}-A_{m}|$ which
causes phase fluctuation, $T_{M}\rightarrow \infty$ as $\ell
\rightarrow \infty$. Therefore $T_{M}$ must have a minimum as a
function of the Fock-Darwin radius $\ell$. At this minimum, $T_{M}\sim
10$ $\mu$s for all $z_{0}$.  For all QD sizes considered here the SD
decay is found to be $\sim \exp{(-\tau^{3})}$, the correlation
function being close to $1$ in the neighborhood of $T_{M}$
(Fig. 9). The dependence with $\theta$ is similar to Fig. 4.


\section{Conclusion}

We have developed a detailed quantitative theory for nuclear spectral
diffusion of localized spins in semiconductors. By treating each
nuclear pair independently, we are able to show that the echo signal
arises from the interplay between fast and slow nuclear flip-flops.
The echo envelope undergoes a smooth transition from a Gaussian SD
decay to an exponential motional narrowing when the nuclear bath loses
correlation over time, this transition being well described by an
appropriate correlation function. The Lorentzian approximation gives a
good description of the intermediate crossover regime, and our theory
gives a microscopic justification for the use of these
phenomenological conditional probabilities.  This behavior is quite
general and should be observed in any decoherence mechanism where
qubit phase fluctuation takes place.  We apply our theory to three
physical systems proposed as QC architectures, showing that SD should
not be a decisive constraint in their development ($T_{M}> 10\:\mu$s;
therefore $T_{M}/\tau_{J}\gg 10^{6}$ for all architectures, where
$\tau_{J}$ is the time scale for the exchange gate\cite{desousa02}).
Our calculated Si:P electron donor spin results agree well with
existing experimental data for natural abundance of $^{29}$Si, while
$T_{M}$ increases very fast as these isotopes are removed from the
lattice.  In addition our calculation shows that the most important
contribution to SD comes from nuclei inside the electron's wave
function spread, as opposed to electrons far away as was suggested
before.\cite{chiba72}  The $^{31}$P nuclear spin is found to be in the
motional narrowing regime, weakly affected by $^{29}$Si. Our GaAs
quantum dot calculations confirm our earlier estimates based on a
simpler theory.\cite{desousa02}  Although there are no Ga or As $I=0$
isotopes, one way to reduce SD is to suppress flip-flop events by
nuclear polarization. The main difference between $T_{1}$ and $T_{2}$
samples is the presence of several relaxation rates. This feature is
evident in Fig. 9, where we show correlation functions for many cases
treated here. Increasing QD radius $\ell$ increases $T_{M}$ and
reduces $\chi_{c}(T_{M})$, pushing large QD's to the motional
narrowing regime. This result is clear evidence that SD does not
affect delocalized states (such as conduction electrons), since all
nuclear pairs will be in the motional narrowing regime [Eq.
(\ref{fastvnm})].
\begin{figure}
\includegraphics[width=3in]{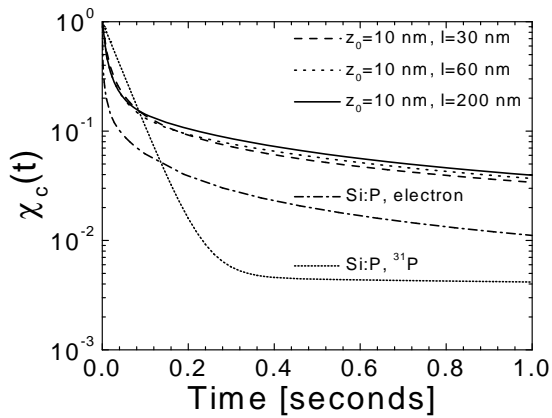}
\caption{  Non-exponential decay of the correlation function
$\chi_{c}$ as a function of time [see
Eq. (\ref{def_correlator})]. Initially $\chi_{c}$ shows strong
exponential decay, dominated by the fastest rate.  Then slower
processes take over. For Si:P we assume natural $^{29}$Si abundance,
while  $B \parallel [111]$. GaAs-QD's show a clear tendency to
motional narrowing as $\ell$ (Fock-Darwin radius) increases.
\label{fignine}}
\end{figure}

We now briefly discuss the approximations assumed in our work.
Certainly the most important one is that we treat nuclear pairs as
independent Markovian processes, making random flip-flops with a
Poisson distribution. We believe this approximation is reasonable for
a nuclear bath at temperatures much higher than its dipolar ordering
critical temperature ($\sim \hbar b_{nm}/k_{B} \sim 10^{-9}$~K), since
in this case the spin correlation length can be assumed of the order
of one lattice parameter. Of course this condition will always hold
for dilution refrigerators (which operate at millikelvin
temperatures), and specially for the experiment discussed in this
paper ($T=1.6$~K).\cite{chiba72} At high spin temperature the nuclear
system undergoes frequent transitions through its $(2I+1)^{N}$
available states. An important assumption implicit in our model is
that only a subset of these transitions, namely flip-flops between
close nuclear pairs [described by $\sigma_{nm}$, Eq.
(\ref{omega_flipflop})] are responsible for the SD effect at external
magnetic fields $B\gg 100$~G.  Four spin processes stem from higher
order perturbations in Eq.  (\ref{fnm}), and can be neglected when
$b_{nm}\lesssim \Delta_{nm}$, which is the case for the relevant
nuclear pairs [otherwise, the condition $b_{nm}\gg \Delta_{nm}$
immediately implies motional narrowing [see Eq. (\ref{fastvnm})] and
this pair gives a negligible contribution to the SD rate].
Preliminary experimental results\cite{tyryshkin03B} on the orientation
dependence of $T_{M}$ show excellent qualitative agreement with Fig.
4, suggesting the pair flip-flop picture is quite appropriate.  One
limitation of the uncorrelated pair approximation is that the echo
peaks always occur at $t=2\tau$.  This happens because the product
[Eq.  (\ref{product})] will be zero away from $2\tau$ as long as a
single pair undergoes slow SD. A global treatment of all pairs such as
the one achieved in the two parameter Gaussian theory leads to an echo
peak formed between $\tau$ and $2\tau$.\cite{klauder62} Nevertheless
to our knowledge this effect does not seem to be present in
$\pi/2-\pi$ echo experiments,\cite{slichter_abragam,chiba72} our
approximation being appropriate in this respect.  Another
approximation that seems reasonable is to neglect back action. We
assume that the nuclei evolve under a static central spin field [Eq.
(\ref{hiprime})], unchanged by its effect on this spin.

Finally, we comment on the relationship between our current work and a
number of other recent publications in the literature dealing with
spin relaxation in the context of semiconductor quantum computer
architectures.\cite{khaetskii01,mozyrsky02,tahan02} We emphasize that
our theory deals exclusively with electron spin decoherence {\em in
  the absence of any phonon effects}. Our work provides a theory for
nuclear-induced spectral diffusion of (localized) electron spin in
semiconductors, and as such we deal with electronic spin decoherence
arising from the fluctuations induced by electron-nucleus hyperfine
interactions caused by nuclear spin dipolar flip-flops. Since the
flip-flops conserve nuclear energy, no phonons are required for the
decoherence studied in this work.

Our work is the first comprehensive theory for nuclear induced
spectral diffusion in the spin decoherence of localized electrons in
semiconductors -- we do not invoke the empirical approximations and
the heuristic arguments which were used to obtain earlier expressions
for spectral diffusion existing in the
literature.\cite{klauder62,chiba72,bloembergen59} As such our results
are applicable and relevant not only to considerations involving solid
state spin-qubit-based quantum computer architectures but also to all
problems involving spin decoherence due to electron-nucleus hyperfine
coupling where phonon effects are negligible (i.e. at low
temperatures).  In particular, our results apply to spin echo
measurements in semiconductors at low temperatures, and it is
therefore gratifying that we have been able to quantitatively explain
hitherto unexplained Si spectral diffusion results of
Ref.~\onlinecite{chiba72} dating back thirty years.  We emphasize that
our theory is still approximate since we are forced to make a number
of approximations, the most important one being the assumption of
uncorrelated flip-flops among spin pairs. Although we believe this
assumption of uncorrelated flip-flops to be well-valid at "high"
nuclear spin temperature ($\gg$ nK), it is still worthwhile to
consider further improvement of our theory by taking into account the
full non-Markovian nature of the spin flip-flop processes in future
work.  The authors acknowledge discussions with S.E. Barrett, A.
Kaminski, S.A. Lyon, J. Fabian, and P. Zoller.  This work is supported
by ARDA, LPS, US-ONR, and NSF.

\appendix
\section{Stochastic theory for a single flip-flopping pair}

The problem is to evaluate the average [Eq. (\ref{cos})] for a single
nuclear pair $n,m$
\begin{equation}
v(t)=\left< \cos{\left[\Delta \int_{0}^{t}S(t')
(-1)^{N(t')}dt'\right]} \right>,\label{cosap}
\end{equation}
with $N(t)$ a Poisson random variable with parameter $t/T$ (for
simplicity we dropped the subscript $nm$ from $v_{nm}$, $\Delta_{nm}$,
$T_{nm}$). Expanding the cosine and rearranging the product of
integrals we get 
\begin{eqnarray}
v(t)&=& \sum_{k=0}^{\infty}
(-1)^{k}\Delta^{2k}\int_{0}^{t}dt_{2k}S(t_{2k}) \nonumber\\
&&\times\int_{0}^{t_{2k}}dt_{2k-1}S(t_{2k-1})\cdots \nonumber\\
&&\times\int_{0}^{t_{2}}dt_{1}S(t_{1})\langle (-1)^{\xi}\rangle,
\label{vtint}
\end{eqnarray}
with $\xi= N(t_{1})+\cdots +N(t_{2k})$.  Using the inequality $t\geq
t_{2k}\geq t_{2k-1}\geq \cdots \geq t_{1}\geq 0$ together with the
fact that a sum of two Poisson variables with parameters $t_{i}/T$ and
$t_{j}/T$ equals another Poisson with $(t_{i}+t_{j})/T$, we get 
\begin{eqnarray}
\xi &=& N(t_{2k}-t_{2k-1})+N(t_{2k-2}-t_{2k-3})+\cdots \nonumber\\
&&+N(t_{2}-t_{1})+2[N(t_{2k-1}) \nonumber\\
&&+N(t_{2k-3})+\cdots+N(t_{1})],
\label{sumn}
\end{eqnarray}
\begin{eqnarray}
\langle (-1)^{\xi}\rangle &=& \exp{\biggr\{
-\frac{2}{T}\left[(t_{2k}-t_{2k-1})+(t_{2k-2}-t_{2k-3})\right.}  \nonumber \\
&&\left.+\cdots+(t_{2}-t_{1})\right]\biggr\},
\label{avgxi}
\end{eqnarray}
since the number of flip-flops $N(t_{2k}-t_{2k-1})$ are assumed
independent random variables for non-overlapping time intervals
$t_{2k}-t_{2k-1}$ (Markovian approximation).  Using Eqs. (\ref{vtint})
and (\ref{avgxi}) we rewrite the echo decay  in the form 
\begin{equation}
v(t)=\sum_{k=0}^{\infty}(-1)^{k}\Delta^{2k}v_{2k}(t)
\label{seriesvt}
\end{equation}
with $v_{2k}(t)$ satisfying the integral recurrence relation
\begin{eqnarray}
v_{2k}(t)&=&
\int_{0}^{t}dt_{2k}\exp{\left(-2\frac{t_{2k}}{T}\right)}S(t_{2k})
\nonumber\\
&&\times\int_{0}^{t_{2k}}dt_{2k-1}\exp{\left(2\frac{t_{2k-1}}{T}\right)}S(t_{2k-1})
\nonumber\\ &&\times v_{2k-2}(t_{2k-1}),
\label{intrec}
\end{eqnarray}
and $v_{0}(t)=1$.  We can transform Eq. (\ref{intrec}) into an
algebraic recurrence relation by using the properties of Laplace
transforms,\cite{arfken95}
\begin{equation}
\tilde{v}_{2k}(p)= \int_{0}^{\infty}e^{-pt}v_{2k}(t)dt.
\end{equation}
The free induction signal, $v_{F}(t)$ [Eq. (\ref{vnmf})] is obtained
by setting $S(t)=1$ for all times. Then Eq. (\ref{intrec}) becomes
\begin{equation}
\tilde{v}_{2k}(p)=\frac{1}{p(p+2T^{-1})}\tilde{v}_{2k-2}(p),
\end{equation}
and the Laplace transform of $v_{F}(t)$ is easily calculated to be 
\begin{equation}
\tilde{v}_{F}(p)= \frac{p+2T^{-1}}{p(p+2T^{-1})+\Delta^{2}}.
\end{equation}
The inverse transform of this expression can be obtained by expanding
in partial fractions,\cite{arfken95}
\begin{equation}
v_{F}(t)=\exp{\left(
-\frac{t}{T}\right)}\left[\frac{1}{RT}\sinh{(Rt)}+\cosh{(Rt)}\right],
\label{vfappendix}
\end{equation}
where $R^{2}=T^{-2}-\Delta^{2}$.  We now turn to the $\pi/2-\pi$ echo
$v_{E}(t)$, which is obtained by setting the echo function to 
\begin{equation}
S(t)=1-2\theta(t-\tau).
\end{equation}
Then Eq. (\ref{intrec}) becomes
\begin{eqnarray}
\tilde{v}_{2k}(p)&=& \frac{1}{p(p+2T^{-1})}\biggl[
\tilde{v}_{2k-2}(p)-2\exp{\left(-p\tau\right)}\nonumber\\
&&\times\exp{\left(-2\tau T^{-1}\right)}
\int_{0}^{\tau}dt'\exp{\left(2t' T^{-1}\right)}\nonumber\\ &&\times
v_{2k-2}(t')\biggl]
\end{eqnarray}
and after summing the series we get 
\begin{eqnarray}
\tilde{v}(p)&=&\frac{p+2T^{-1}}{\Delta^{2}+p(p+2T^{-1})}\nonumber\\
&&+\frac{2\Delta^{2}\exp{(-p\tau)}}{\Delta^{2}+p(p+2T^{-1})}f(\tau),\\
f(\tau)&=&\exp{\left( -2\tau T^{-1}\right)}
\nonumber \\
&&\times\int_{0}^{\tau}dt'
\exp{\left( 2t'T^{-1}\right)}v_{F}(t').
\end{eqnarray}
Using Eq. (\ref{vfappendix}), the inverse transform becomes  
\begin{eqnarray}
v(t)&=&v_{F}(t)+\theta(t-\tau)
\frac{2\Delta^{2}}{R^{2}}\exp{\left(-tT^{-1}\right)} \nonumber\\
&&\times\sinh{(R\tau)}\sinh{\left[R(t-\tau)\right]}\\ &=&
\theta(\tau-t) v_{F}(t)+\theta(t-\tau)v_{E}(t),
\end{eqnarray}
which after rearrangement leads to Eq. (\ref{vnm}) for $t>\tau$,
\begin{eqnarray}
v_{E}(t)&=& R^{-2}\exp{\left(-tT^{-1}\right)}\bigg\{
T^{-2}\cosh{(Rt)}\nonumber\\
&&+RT^{-1}\sinh{(Rt)}\nonumber \\
&&-\Delta^{2}\cosh{\left[ R(t-2\tau)\right]}
\bigg\}.
\label{vnmechoappendix}
\end{eqnarray}

\section{Equivalence of ensemble $\pi/2 -\pi$ echo envelope and 
average single spin dynamics}

Eq. (\ref{cos}) was derived by averaging $\sigma_{nm}(t=0)$, which
clearly applies only to an ensemble of spins. Therefore an interesting
question is how the single spin off diagonal density matrix element
$\langle S_{\perp}\rangle$ [analogous to Eq. (\ref{vt}), the in-plane
magnetization in the $S_{z}$ basis but without $\sigma_{nm}(0)$
average] behaves under nuclear dynamics average only. The result is
that the modulus squared of this  quantity is exactly equal to the
$\pi/2-\pi$ echo envelope, as we show below. The evolution under
Eq. (\ref{omega_flipflop}) is given by
\begin{equation}
\langle S_{\perp}(t)\rangle = \prod_{n<m}\left< \exp{\left[  i
\Delta_{nm}\int_{0}^{t}(-1)^{N_{nm}(t')}dt'\right]}\right>,
\label{singleoffdiag}
\end{equation}
where $N_{nm}(t)$ are independent Poisson random variables
representing the nuclear dynamics, on which the average is
taken. Using the same methods of Appendix A, it is easy to show that
the argument of the product (\ref{singleoffdiag}) is given by
\begin{eqnarray}
S^{\perp}_{nm}(t)&=&\left< \cos{\left[\Delta_{nm}
\int_{0}^{t}(-1)^{N_{nm}(t')}dt'  \right]}\right> \nonumber \\ &&+
\left<\sin{\left[\Delta_{nm} \int_{0}^{t}(-1)^{N_{nm}(t')}dt'
\right]}\right> \nonumber\\ &=& v_{nm}^{(F)}(t) \nonumber \\ &&+ i
\frac{\Delta_{nm}}{R_{nm}} \exp{\left(
-\frac{t}{T_{nm}}\right)}\sinh{(R_{nm}t)},
\label{snm}
\end{eqnarray}
where $v_{nm}^{(F)}(t)$ is the free induction decay derived above
[Eq. (\ref{vnmf})]. Clearly the difference between a single spin and
an ensemble is the presence of the complex term in Eq. (\ref{snm}),
which leads to strong interference effects when the product over pairs
is taken. To see this, we calculate 
\begin{eqnarray}
|S^{\perp}_{nm}(t)|^{2}&=&  [v_{nm}^{(F)}(t)]^{2} \nonumber \\  &&+
\frac{\Delta_{nm}^{2}}{R_{nm}^{2}} \exp{\left(
-\frac{2t}{T_{nm}}\right)}\sinh^{2}{(R_{nm}t)}\nonumber \\ &=&
v_{nm}^{(E)}(2t).
\end{eqnarray}
Therefore the effect of this complex part is to enhance the coherence
of the single spin, making it exactly equal to the $\pi/2-\pi$ echo
envelope,
\begin{equation}
|\langle S_{\perp}(\tau)\rangle|^{2}=v_{E}(2\tau).
\end{equation}
Notice that if we performed the product over pairs without this complex
part, $S_{\perp}$ would decay similarly to free induction, $v_{F}(t)$.
Many authors define a coherence time $\tau_{c}$ equal to the $1/e$
decay of the modulus of the off diagonal density matrix. If the echo
is dominated by Gaussian SD, we simply have
$\tau_{c}=T_{M}/\sqrt[3]{4}$.

\end{document}